# Modeling Superionic Behavior of Plutonium Dioxide

S.D. Günay[a,*], B. Akgenç[a,b], Ç. Taşseven[a]

[a]Yıldız Technical University, Department of Physics, Faculty of Science, Esenler, 34210, Istanbul,Turkey

[b]Kırklareli University, Department of Physics, Faculty of Science, Kavaklı, 39060, Kırklareli, Turkey

The Bredig transition to the superionic phase indicated with the $\lambda$-peak in $C_p$ was highly expected for $PuO_2$ as other actinide dioxides. However, least-square fit and local smoothing techniques applied to the experimental enthalpy data of plutonium dioxide in 80's could not detect a $\lambda$-peak in specific heat that might be due to too scattered and insufficient experimental data. Therefore, this issue has not been yet put beyond the doubts. In the current article, a superionic model of $PuO_2$ is developed with partially ionic model of a rigid ion potential. Thermophysical properties were calculated in constant pressure-temperature ensemble using molecular dynamics simulation. The Bredig transition with vicinity of a $\lambda$-peak in specific heat was a successfully observed for the model system at about 2100K. Moreover, the experimental enthalpy change was well reproduced before and after the estimated transition temperature.

---

[*] Corresponding Author. Address: Yıldız Technical University, Department of Physics, Faculty of Science, Esenler, 34210, Istanbul, Turkey. Tel./fax: +90 212 3834289.
E-mail address: sdgunay@gmail.com (S. D. Günay).

## 1. Introduction

Plutonium dioxide $\left(PuO_2\right)$, generally used in mixed compound (MOX), is an important material as a fuel and a stable storage instrument in nuclear reactors. Our knowledge of this material is limited due to toxicity and radioactivity which make it difficult to study. The melting temperature of $PuO_2$ is an important parameter in studying phase diagram of its mixed oxide, generally with $UO_2$, at high temperature. There have been a few experiments that aim to determine the melting temperature of $PuO_2$ [1,2,3] which was apprised to be around 2700K. However, during the traditional heating techniques, the measurements were affected either by the reaction of oxygen sample with the metallic containment or by the loss of oxygen in the atmosphere [4,5,6].

Recently, the melting behavior of stoichiometric $PuO_2$ has been studied for the first time by the fast laser heating and multi-wavelength spectro-pyrometry to reduce the side effect mentioned above [6]. Based on this measurement the melting point of $PuO_2$ has been determined as $3017 \pm 28K$. Moreover, the temperature evaluation of physicochemical properties of $PuO_2$ are limited up to 2000K which is well below the melting temperature [7,8]. Even though, a diffuse transition called Bredig transition is a highly expected feature of all the actinide oxide yet it has not been proved experimentally for $PuO_2$. In order to observe whether $PuO_2$ display the Bredig transition to the superionic phase or not one has to produce some physicochemical properties up to melting point. However, during the enthalpy measurements of $PuO_2$ by Ogard [7], partial melting of sample above 2370K was observed that has been thought, from the private communication [7], as a result of the reaction between the container and $PuO_2$. Therefore, Fink did not include the phase transition in the non-linear least-square fit to the experimental data due to the large scatter data points between 2160-2370K and lack of data above 2370K. Later on, Ralph [8] has applied the method of quasi-local linear regression (QLLR) to the enthalpy data by Ogard and no peak has been produced in both the real $C_p$ and the artificial $C_p$. It is interpreted that the absence of the $C_p$ peaks may be the result of the systematic error in the regression together with the random error in the data.

At this point, molecular dynamics simulation turns out to be a useful tool for studying the properties which may be essential for nuclear facilities to run safely. Molecular dynamics simulation of such systems requires developing a reliable potential for the interactions. Within the nuclear fuel materials, uranium dioxide $UO_2$ is one of the most studied systems. There have been a number of pair potentials developed to understand thermophysical and transport properties of $UO_2$ at solid and liquid phases. Govers et al [9,10] have made a broad comparison to assess the range of applicability of these potentials. For plutonium dioxide, previous studies have come up with some pair potentials, generally in the form of Buckhingham, Morse, embedded atom model (EAM) and shell model to investigate $PuO_2$ [11-15]. Recently, Cooper et.al. [15] have reported a many-body potential model to describe the thermomechanical properties of actinide oxides between 300K and 3000K. However, the Bredig transition has not been considered at any stage of this work.

Plutonium dioxide, like other actinide dioxides, is believed to be a type-II superionic conductor which displays a rapid but continuous increase in the ionic conductivity by heating at about $0.8T_m$. The extent of anion disorder increases within the superionic phase. Ionic conductivity occurs almost entirely with anion diffusion. Although the phase transition is well above the operating temperature of nuclear facilities, anion frenkel pair concentration begins to increase at this level. So the detailed knowledge of the Bredig transition and the melting of $PuO_2$ at atomic level is important in order to enhance the nuclear fuel performance and the accident conditions.

The latest value of the melting temperature of $PuO_2$ determined by laser-heating technique gave us the opportunity to investigate the thermally induced Bredig transition and the melting. For this purpose, with the present article, a new semi-empirical rigid ion potential is introduced and thermo-physical properties are calculated via classical molecular dynamics simulation in constant pressure-temperature ensemble are presented for the wide range of temperature (300K-3600K).

## 2. Potential model

There exist a few semi-empirical potential model in literature developed for $PuO_2$ [11-15]. Yamada, Kurosaki and Arima [11-13] have used Morse and Buckhingham type pair interaction potential function. Buckhingham type with shell model pair interaction potential has been introduced by Chu [14]. Cooper added the term many body EAM to Buckhingham and Morse type pair interaction potentials.

In general, the parameters for these potential functions O-O interaction have been generally taken from the potential developed earlier for $UO_2$. On the other side, Pu-Pu and Pu-O parameters have been determined by try and error method aiming to reproduce the experimental lattice constant at low temperatures. In obtaining the parameters of the pair potential in the present study, we also take the O-O parameters from the potential developed for $UO_2$ in our previous work [16]. The other parameters were adjusted in order to reproduce low temperature experimental values of lattice parameter, bulk modulus, elastic constants and cohesive energy. Moreover, we have also aimed to model the Bredig transition to the superionic phase which is highly expected but has not been yet observed experimentally. The potential function has been originally proposed by Vashishta-Rahman [17] and is given by,

$$\phi_{ij}(r) = \frac{q_i q_j}{r} + \frac{A_{ij}\left(\sigma_i + \sigma_j\right)^{\eta_{ij}}}{r^{\eta_{ij}}} - \frac{P_{ij}}{r^4} - \frac{C_{ij}}{r^6} \tag{1}$$

The first term stands for the Coulomb interactions, second term contains core repulsions, third term is the effective monopol-induced dipole interaction and the last term is Van der Waals interaction. The potential parameters are given in Table 1.

### 3. Molecular dynamics simulation

Molecular dynamics technique were performed for 5x5x5 cell constructed with 500 cations ($Pu^{4+}$) and 1000 anions ($O^{2-}$) arranged as $CaF_2$ type structure. The MD code called as MOLDY [18] was adopted to carry out the calculations. The Ewald's sum technique is used to account for the long range Coulomb interactions. The positions and velocities of the ions are calculated by integrating the Newton's equation of motion using Beeman's algorithm, which is predictor-corrector type, with the time step $\Delta t = 1.0x10^{-15} s$. Nose-Hoover thermostat and Parrinello-Rahman constant pressure methods are applied to control the temperature and pressure in the constant pressure-temperature (NPT) ensemble. The simulations were run for 50 ps: the first 10 ps was used for the equilibration process and the properties of the system were calculated by averaging over the following 40 ps. The temperature is varied from 300 K to 3600 K at 100 K intervals. Additional calculations have been carried out with different number of atoms each for 50K temperature intervals: 864 ($Pu^{4+}$)-1728 ($O^{2-}$), 1372 ($Pu^{4+}$)-2744 ($O^{2-}$) and 2048 ($Pu^{4+}$)-4096 ($O^{2-}$); 6x6x6, 7x7x7 and 8x8x8, respectively, to make sure that number of atoms are enough to calculate thermodynamic properties and also to be able to determine the type of the transitions. In order to calculate properties, these simulations have been performed for 100ps where the equilibration run is 30ps and data is accumulated for 70ps.

### 4. Results and discussion

The values of lattice parameter, bulk modulus, elastic constants and cohesive energy are given in table 2 and compared with the available experimental data and ab-initio results. The bulk modulus and elastic constants were calculated from the utilization of known Birch-Murnaghan equation of state [19, 20] which is in good agreement with experiment and ab-initio results [21-25]. Contrary to the previous MD results [11-15], in the present paper the lattice parameter is underestimated about 1.3% and 3% compared to the ab-initio calculations and experimental data, respectively [22, 23]. However, bulk modulus obtained with the previous potentials has been over estimated and varied between 200-239 GPa [11-15] when compared with experimental result 178 Gpa [21]. In the present study, the amount of deviation of bulk modulus from the experimental study is about 5% whereas other studies are between 12%-34%. The main reason for these results is the weight of observables while finding the potential parameters. Groups place great emphasis on the lattice constant while developing potential parameters. In this study, we distribute the weight constant unevenly on lattice constant, bulk modulus, elastic constants and energy as we are performing fitting procedure. Temperature dependence of energy is also taken into account in order to mimic the fast-ion phase transition.

$L/L_0$ is calculated where $L_0$ is the lattice constant at 300K, as a function of temperature together with the other simulation results [11-15] and the experimental data [22] which are available up to 2000 K are presented in Figure 1.

Figure 2 shows the linear thermal expansion $\Delta L/L_0$ . All potentials produce a similar thermal expansion of the lattice between 300K and 2000K; discrepancy between the simulation and the experiment increases as the temperature increased. At about 2100 K, there is a sign of a change in trend of the evaluation of the data with temperature. As recommended by Fink [7] transition to the superionic phase may be expected at temperature between 2160-2370 K.

The temperature dependence of the pair correlation functions $g_{ij}(r)$ given in Figure 3 were also considered in our calculations aiming to have quantitative measure of the spatial local order-disorder of the atomic structure of $PuO_2$.

The radial distribution function is defined as following,

$$g(r) = \frac{\Delta n(r)}{\rho 4\pi r^2 \Delta r} \qquad (2)$$

$\Delta n(r)$ is the average number of atoms at a distance between $r$ and $r + \Delta r$. $\rho$ is the atom density and $\Delta r$ is the width of the shell. The peaks of $g_{ij}(r)$ of all three pairs of ions become lower and broader due to the larger thermal vibrations of ions in their lattice sites as the temperature increased up to 2000K. Right after 2000K, $g_{oo}(r)$ has very small oscillations and overlap of the principal peaks after the first peak indicating a considerable degree of disorder of the oxygen sublattice while the plutonium ions are still located in the lattice sites with larger vibrations. This liquid like behavior of the oxygen sublattice at solid phase at about 2100K is interpreted as the onset of the transition to the superionic phase of $PuO_2$.

For a species of N particle, the mean square displacement (MSD) is calculated as

$$\left\langle \left| r(t) - r(0) \right|^2 \right\rangle = \frac{1}{NN_t} \sum_{n=1}^{N} \sum_{t_0}^{N_t} \left| r_n(t + t_0) - r_n(t_0) \right|^2 \qquad (3)$$

where $r_n(t)$ is the position of particle $n$ at time $t$. The mean square displacement (MSD) of oxygen ions at different temperatures are shown in Figure 4. The linear change of MSD with time indicates that the oxygen ions become more diffusive at about 2100K while the MSD of Pu ions remains constant. Figure 5 illustrates the diffusion coefficient of both Pu and O ions in Arrhenius diagram (log D vs. 1/T) that are calculated from the slope of the MSD data using Einstein relation,

$$D_i = \lim_{t \to \infty} \frac{1}{6t} \left\langle \left| r_i(t) - r_i(0) \right|^2 \right\rangle \qquad (4)$$

It is clearly evident from the anomalous increase of diffusion coefficient of oxygen ions at ~2100 K that there is an onset of Bredig transition to the superionic phase. While the oxygen diffusion continuously increases beyond 2100 K, the plutonium ions show almost no diffusive behavior. As experimental data for oxygen diffusion coefficient for $PuO_2$ is only available up to 1400K [26], we are not able to compare self-diffusion coefficient. As the temperature increases Pu ions also show diffusive behavior after 3000K and $D_{Pu}$ is less than $D_O$ about order of 10.

The enthalpy change $\Delta H = H_T - H_{298}$ is presented in Figure 6 and compared with the experimental data taken from Fink [7] and MD simulation with shell model potential by Chu. Evidently, there is a discontinuity in calculated $\Delta H$ at about 2100K and very good agreement with experimental data before and after this temperature. This justifies the potential parameters and validates the constructed model of $PuO_2$.

The existence of the thermally activated transition into superionic phase can be confirmed by a λ-peak in the heat capacity at constant pressure $C_p(T)$. Many fluorite type ionic crystals exhibit such a transition at about $0.8T_m$ [27]. The heat capacity was evaluated from the variation of the internal energy $E(T)$ with temperature at constant pressure and presented in Figure 7.

$$C_p(T) = \left(\frac{\partial E}{\partial T}\right)_p \qquad (4)$$

A very weak increase in Cp up to 1750 K is interpreted as increase in the anharmonicity of the lattice vibrations. Further increase in temperature results with the additional increase in calculated Cp and a λ shape peak is clearly produced at the critical temperature Tc=2055 K which is close to the expected value of the phase transition temperature in fluorite type of ionic crystals [7]. In order to make sure that it is a λ -peak transition in specific heat, we have increased the number of data by varying the temperature at 50K intervals in NPT-ensemble for MD simulation calculations. These calculations have been carried out for three different supercell boxes: 6x6x6 (864 Pu-1728 O), 7x7x7 (1372 Pu-2744 O) and 8x8x8 (2048 Pu-4096 O) to show the consistency of the calculations in term of number of atoms. Results for heat capacity at constant pressure around the transition region are shown in Figure 7 as inset. It is clearly evident that the system undergoes a thermally activated λ-type Bredig transition to the superionic phase. Moreover, the linear increase of mean square displacement of oxygen ions with time at about 2100K while that of plutonium remains constant presented in Figure 4 and anomalous increase of diffusion coefficient of oxygen ions at around the same temperature presented in Figure 5, that are due to the premelting of oxygen sublattice, also support the indication of lambda transition in heat capacity.

## 5. Conclusions

A model of plutonium dioxide $\left(PuO_2\right)$ as a superionic conductor has been constructed via parameterization of a rigid ion potential with partially ionized atoms. The thermophysical properties of the model have been investigated up to 3600K within the classical molecular dynamics in the NPT ensemble. A sign of an anomalous behavior in the lattice parameter and oxygen-oxygen pair correlation function was observed at about 2100K. This was reflected as a sharp increase in the self oxygen diffusion coefficient to the value of a typical liquid system, a discontinuity in enthalpy change and as a λ-type peak in the constant pressure heat capacity at about the same temperature. These features indicate that the system undergoes a thermally activated Bredig transition to the superionic phase and validates the constructed superionic model of $PuO_2$. As it is also pointed out by Fink [7] more accurate high-temperature enthalpy measurements are needed to put this issue beyond the doubts.

## Acknowledgements

This research has been supported by Yıldız Technical University Scientific Research Projects Coordination Department. Project Number: 2013-01-01-GEP01.

**Table 1** The potential parameters used in the present work. $Z_{Pu} = 1.96e$ and $Z_O = -0.98e$.

**Table 2** Comparison of calculated results with MD simulation, experimental and ab initio data.

**Figure 1.** Lattice parameter evolution with temperature. Experimental data are from TPRC data series [22].

**Figure 2.** Evolution of relative linear thermal expansion of present results, MD data and experimental data with temperature.

**Figure 3.** Pair correlation functions of $PuO_2$ at three different temperatures.

**Figure 4.** Mean square displacement versus time at different temperatures.

**Figure 5.** Evolution of diffusion coefficient with temperature for both Plutonium and Oxygen ions.

**Figure 6.** Enthalpy change of $PuO_2$ with temperature.

**Figure 7.** Temperature dependence of heat capacity for $PuO_2$ for the present study at 100K intervals and comparison with experimental data. The inset shows the results for the different box sizes at 50K intervals.

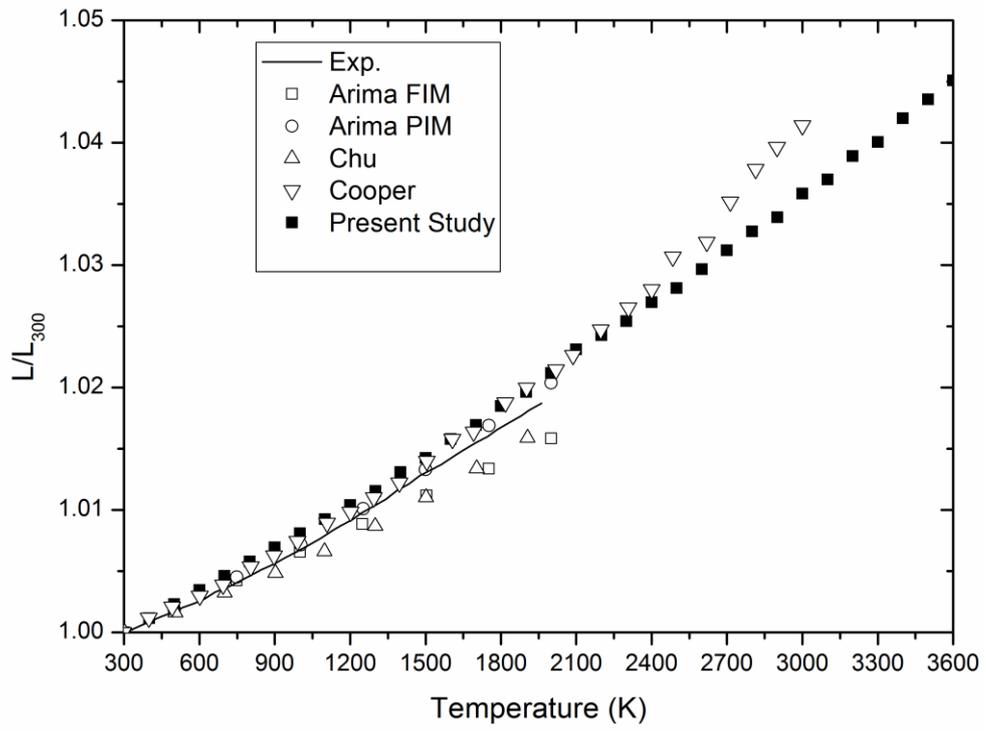

Figure 1.

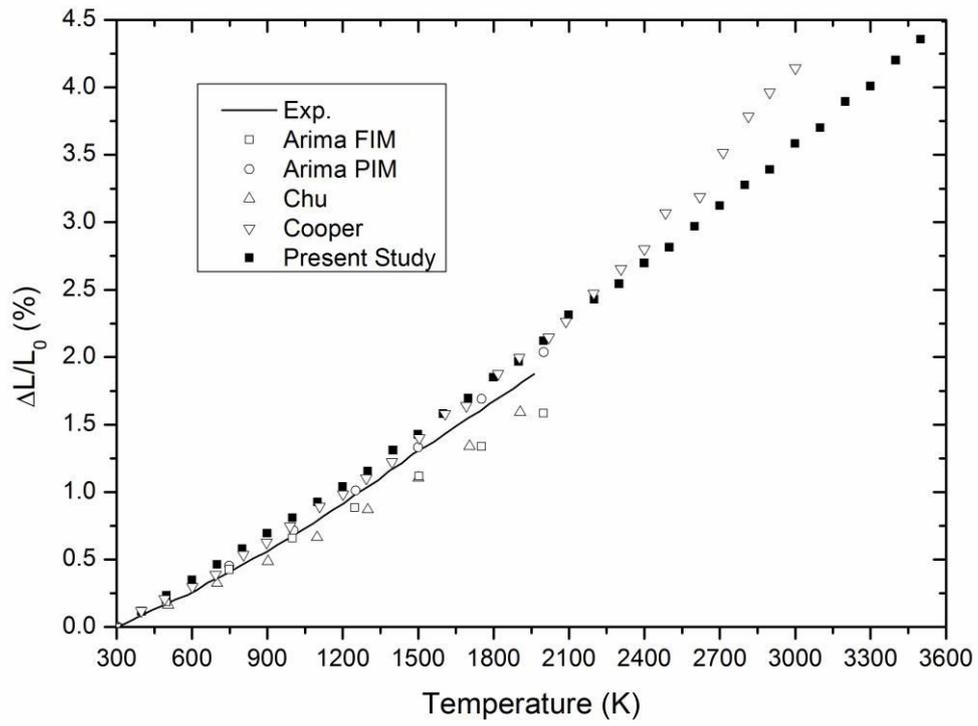

Figure 2.

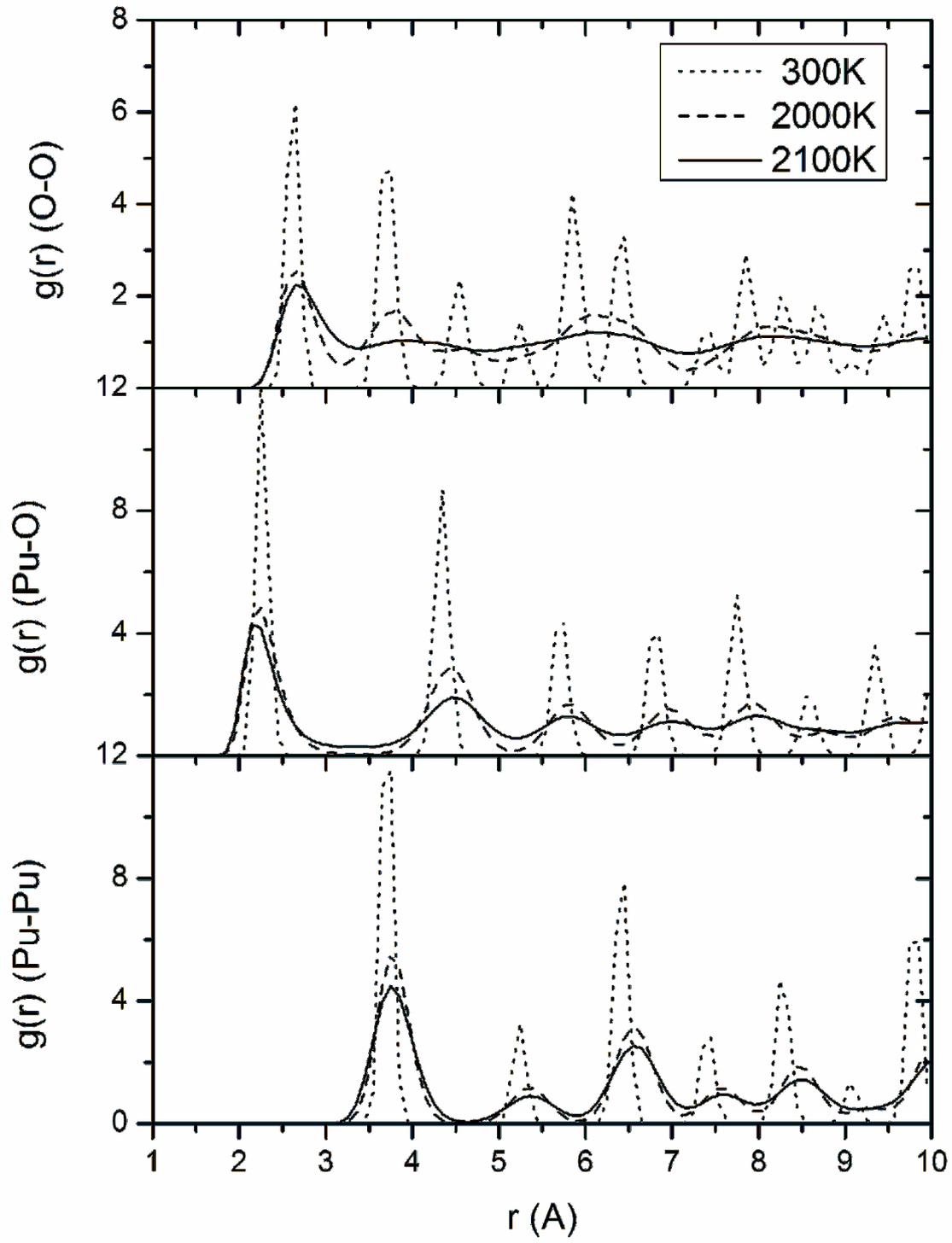

Figure 3.

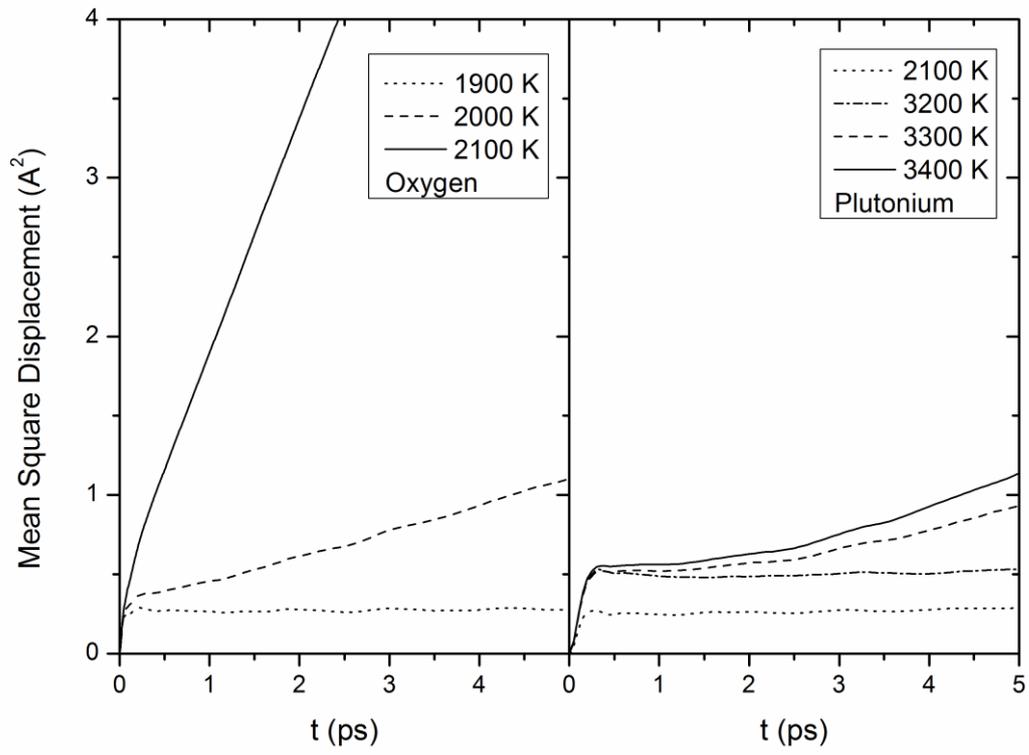

Figure 4.

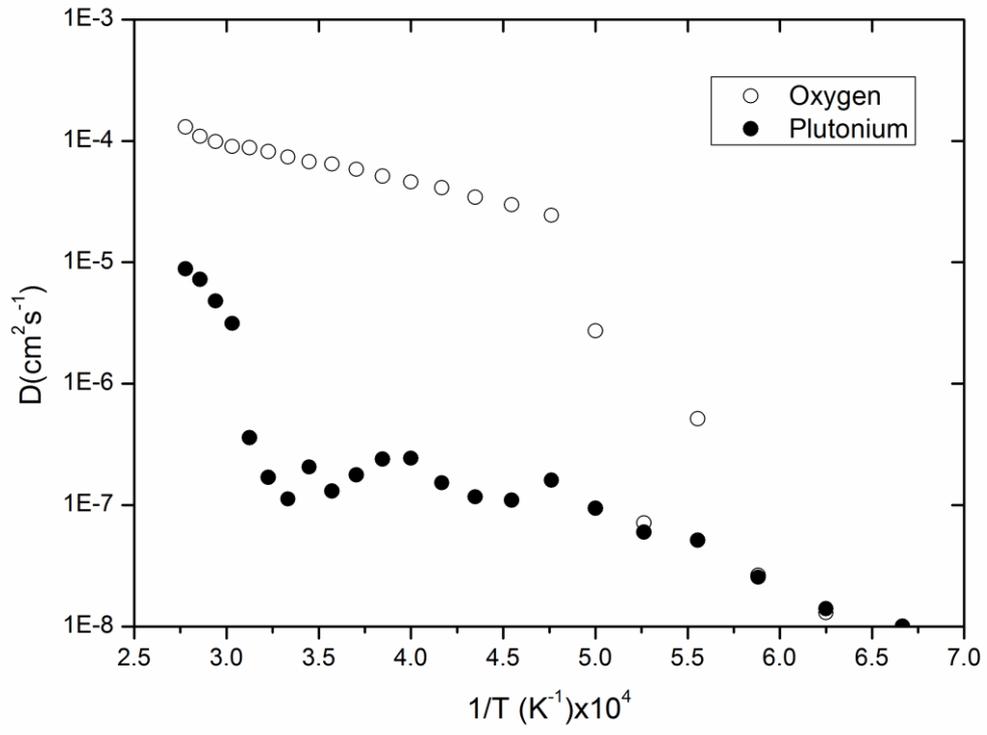

Figure 5.

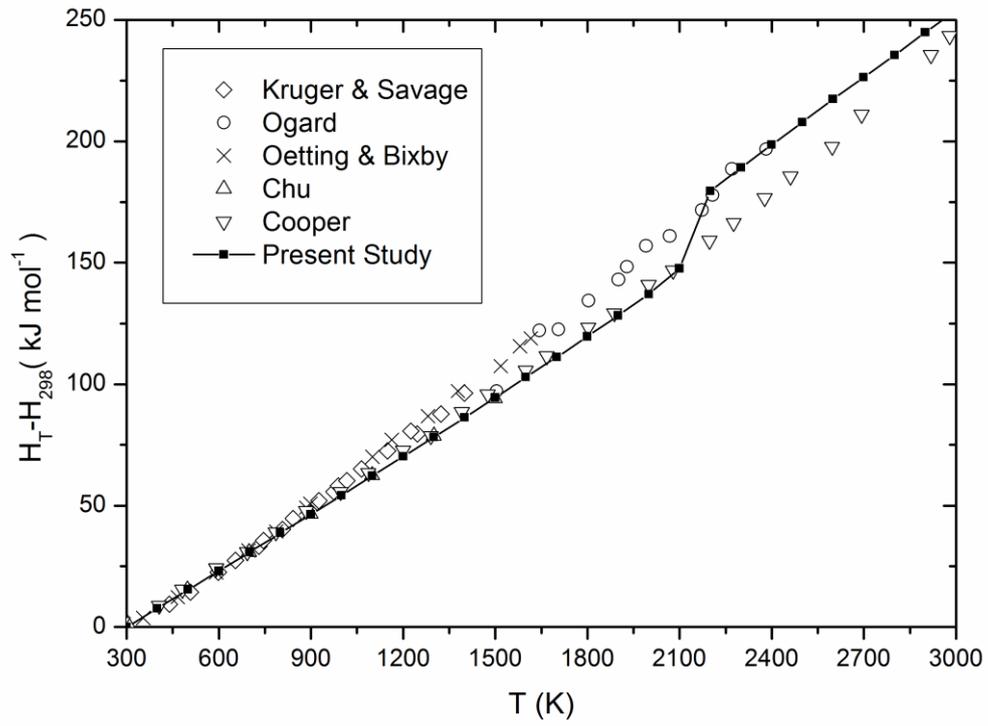

Figure 6

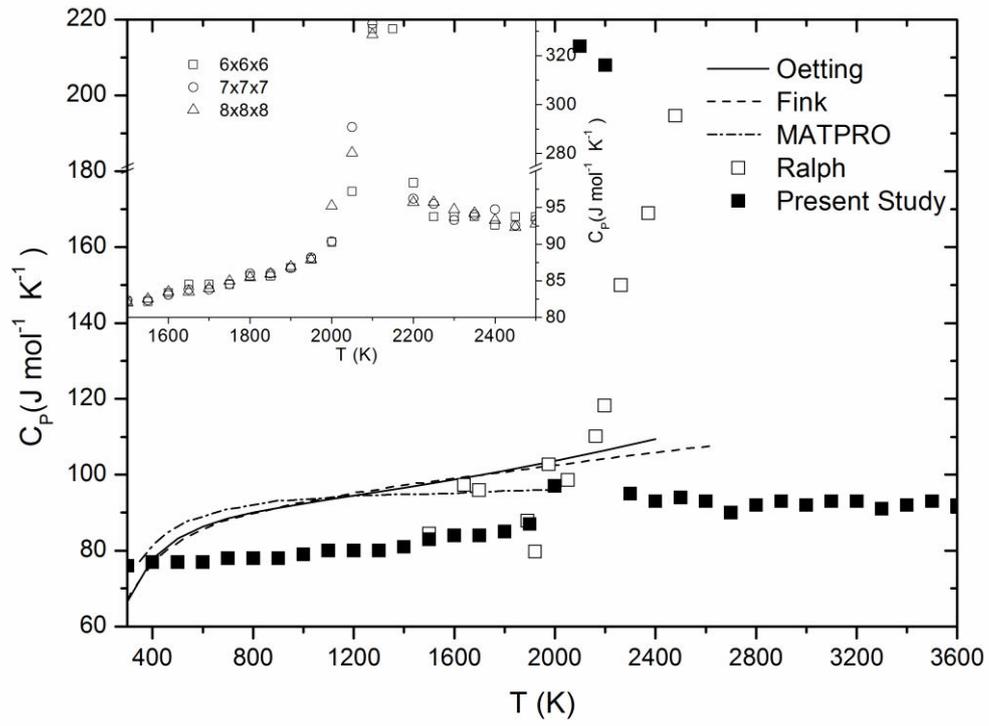

Figure 7.

**Table 1.**

|  | $A_{ij}$ (eV) | $P_{ij}$ (eV Å$^4$) | $C_{ij}$ (eV Å$^6$) | $\eta_{ij}$ | $\sigma_{i,j}$ (Å) |
|---|---|---|---|---|---|
| Pu-Pu | 1.52658 | 0.0 | 0.0 | 7 | 1.37 |
| Pu-O | 0.5238 | 0.0 | 0.0 | 7 | |
| O-O | 4.0859 | 40 | 8.3 | 7 | 1.00 |

**Table 2**

|  | Present | Exp. / Ab initio | MD Simulation |
|---|---|---|---|
| $B_0$ (GPa) | 168 | 178 [21] | 200-239 [11-15] |
| $a_0$ (Å) | 5.235 | 5.307-5.398 [22,23] | 5.389-5.39 [11-15] |
| $C_{11}$ (GPa) | 325 | 256-386 [24] | 424.3 [15] |
| $C_{12}$ (GPa) | 88 | 112-177 [24] | 111.7 [15] |
| $C_{44}$ (GPa) | 74 | 53-74 [24] | 69.2 [15] |
| $E_C$ (eV/PuO$_2$) | 31.8 | 19.24-24 [23,25] | - |